\documentclass[axioms,article,accept,pdftex,moreauthors]{Definitions/mdpi} 

\firstpage{1} 
\pubvolume{1}
\issuenum{1}
\articlenumber{0}
\pubyear{2024}
\copyrightyear{2024}
\externaleditor{Academic Editor: }
\datereceived{29 November 2023} 
\daterevised{21 January 2024} 
\dateaccepted{23 January 2024} 
\datepublished{ } 
\hreflink{https://doi.org/} 

\usepackage{rotating}

\Title{Computing Transiting Exoplanet Parameters with 1D Convolutional Neural Networks}

\TitleCitation{Computing Transiting Exoplanet Parameters with 1D Convolutional Neural Networks}

\Author{Santiago Iglesias Álvarez 
$^{1,}$*\orcidA{}, Enrique Díez Alonso $^{1,2,}$*\orcidC{}
, María Luisa Sánchez Rodríguez $^{1,3}$\orcidB{}, Javier~Rodríguez~Rodríguez $^{1}$\orcidD{}, Saúl Pérez Fernández $^{1}$ \orcidE{} and Francisco Javier de Cos Juez $^{1,4}$\orcidF{}}

\AuthorNames{Santiago Iglesias Álvarez, Enrique Díez Alonso, María Luisa Sánchez Rodríguez, Javier Rodríguez Rodríguez, Saúl Pérez Fernández and Francisco Javier de Cos Juez}

\AuthorCitation{Iglesias Álvarez, S.; Díez Alonso, E.; Sánchez Rodríguez, M.L.; Rodríguez Rodríguez, J.; Pérez Fernández, S.; de Cos Juez, F.J.}

\address{%
$^{1}$ \quad Instituto Universitario de Ciencias y Tecnologías Espaciales de Asturias (ICTEA), C. Independencia 13, \mbox{33004 Oviedo, Spain}; mlsr@uniovi.es (M.L.S.R.); rodriguezrjavier@uniovi.es~(J.R.R.); perezsaul@uniovi.es~(S.P.F.); fjcos@uniovi.es (F.J.d.C.J.)\\
$^{2}$ \quad Departamento de Matemáticas, Facultad de Ciencias, Universidad de Oviedo, 33007 Oviedo, Spain\\
$^{3}$ \quad Departamento de Física, Universidad de Oviedo, 33007 Oviedo, Spain\\
$^{4}$ \quad Departamento de Explotación y Prospección Minera, Universidad de Oviedo, 33004 Oviedo, Spain\\}

\corres{Correspondence: iglesiassantiago@uniovi.es (S.I.Á.);
diezenrique@uniovi.es (E.D.A.)}

\abstract{The transit method allows the detection and characterization of planetary systems by analyzing stellar light curves. Convolutional neural networks appear to offer a viable solution for automating these analyses. In this research, two 1D convolutional neural network models, which work with simulated light curves in which transit-like signals were injected, are presented. One model operates on complete light curves and estimates the orbital period, and the other one operates on phase-folded light curves and estimates the semimajor axis of the orbit and the square of the planet-to-star radius ratio. Both models were tested on real data from TESS light curves with confirmed planets to ensure that they are able to work with real data. The results obtained show that 1D CNNs are able to characterize transiting exoplanets from their host star's detrended light curve and, furthermore, reducing both the required time and computational costs compared with the current detection and characterization algorithms.}

\keyword{astrophysics; transits; exoplanets; neural networks; convolutional neural networks; artificial intelligence; simulations}

\MSC{85-00; 85A99; 68T07; 68T10; 85-10}

\begin{document}

\section{Introduction}
Exoplanet detection is one of the most relevant fields in astrophysics nowadays. Its origins trace back to 1992, when the discovery of three exoplanets orbiting the $PSR 1257+12$ pulsar \citep{1992Natur.355..145W} emerged from data collected by the 305 m Arecibo radio telescope. This discovery was the beginning of a new research area that is still being studied nowadays, although this planetary system is not what they were looking for, as~the surroundings of a pulsar are completely different from the one surrounding a star in which there could be planets similar to the~Earth.

Over the years, various detection techniques have been established. One of the most used is the transit method, which consists of detecting periodic dimmings in the stellar light curves (which are the flux in the function of the time) due to the crossing of an exoplanet (or more) with the line of sight between its host star and a telescope monitoring it. This is probably the most used technique nowadays, as~there are a lot of photometry data available from different surveys, as~one telescope can monitor thousands of stars at the same time. The~first discovery of exoplanets related to this technique took place in the year 2000, when~\cite{Charbonneau_2000, Henry_2000} discovered an exoplanet orbiting the star HD 209458. The~dimmings detected through this technique are described by the Mandel and Agol theoretical shape~\cite{Mandel_2002}, which takes into account an optical effect known as limb darkening, which makes the star appear less bright at the edges than at the center. From~these models, it is possible to estimate the orbital period (P), which is the temporal distance between two consecutive transits; the planet-to-star radius ratio ($R_p/R_\star\equiv R_p[R_\star]$), which is related to the transit depth; the semimajor axis of the orbit in terms of the stellar radius ($a/R_\star\equiv a[R_\star]$); and the orbital plane inclination angle (i).

The main challenges related to the analysis of light curves when trying to search and characterize transit-like signals are the high computational cost required to analyze the large dataset of light curves available from different surveys, the~high amount of time required to visually inspect them; and the fact that stellar noise present in light curves could make the transit~detection considerably more difficult.

One of the most relevant solutions to these problems comes from artificial intelligence (AI). If~an AI model could be able to distinguish between transit-like signals and noise, it would reduce the amount of light curves that are needed to analyze with current~algorithms.

Current algorithms, such as box least squares (BLS) \cite{BLS} and transit least squares (TLS)~\cite{TLS}, among~others, could be classified mainly into Markov chain Monte Carlo (MCMC) methods and least squares algorithms. The~second ones, including BLS and TLS, search for periodic transit-like signals in the light curves and compute some of the most relevant parameters related to the transit method (as P, $ R_p[R_\star]$, etc.). However, BLS needs about 3 s per light curve and TLS about 10 s (taking into account the simulated light curves explained in {Section \ref{Sec: lc_sim}} and the server used in this research (see {Section \ref{sec: train}})), making the analysis of thousands of light curves within a short timeframe exceedingly~challenging.

The TESS (Transiting Exoplanet Survey Satellite) mission~\cite{2015JATIS...1a4003R} is a space telescope launched in April 2018, whose main goal is to discover exoplanets smaller than Neptune through the transit method, orbiting stars bright enough to be able to estimate their companions' atmospheres. The~telescope is composed of 4 cameras with 7 lenses, thus allowing for monitoring a region of $24^\circ \times \:96^\circ$ during a sector (27 days). It took data with long and short cadence (30 min and 2 min, respectively). Its prime mission started in July 2018 and finished in April 2020. Currently, it is developing an extended mission, which started in May 2020. It is important to remark that TESS light curves obtained from full frame images (FFIs) usually present high noise levels in less bright stars, which makes planetary detection and~characterization considerably more difficult.

Understanding exoplanet demographics in the function of the main stellar parameters is important not only for checking and improving planetary evolution and formation models, but~also for studying planetary habitability~\cite{2021exbi.book....2G}. The~main parameters on which to calculate demographics can be split into three main groups: planetary system, host star, and surrounding environment. First of all, it is important to remark that all detection techniques have clear bias related to which type of planets are able to detect, which greatly conditions the study of planetary demography. One example is the radial velocity (RV) technique, which consists of measure Doppler shifts in stellar spectra due to the gravitational interaction between the planets and their host star, which is highly dependent on the planet-to-star mass ratio. From~\cite{1999ApJ...526..890C, 2008PASP..120..531C}, it was estimated that $\sim$20\% of solar-type stars host a giant exoplanet (i.e., $M_P > 0.3~M_J$, where $M_J$ is the Jupiter mass) at 20 AU. The~main contributions of RV research to planet demographics (\cite{2005ApJ...622.1102F, 2010PASP..122..905J} among others) include that, in~solar-type stars and considering a short orbital period regime, low mass planets are up to an order of magnitude more probable than giant ones. In~addition, the~giant planet occurrence rate increases along with the period up to about 3 AU from its host star, and~also, if~they are closer to 2.5 AU, these rates increase with stellar mass and metallicity ($[Fe/H]$). Finally, Neptune-like planets are the most frequent beyond the frost line~\cite{Suzuki_2016} (the minimum distance from the central protostar in a solar nebula, where the temperature drops sufficiently to allow the presence of volatile compounds like water). Other techniques, such as transits, also help to study planetary demographics. It is necessary to highlight the contributions of the Kepler~\cite{2010Sci...327..977B} mission. Its exoplanet data allowed the study of the bimodality in the small planet size's distribution~\cite{2018AJ....156..264F} and the discovery of the Neptune dessert~\cite{Mazeh}, which is a term used to describe the zone near a star ($P<2-4d$) where the absence of Neptune-sized exoplanets is observed. In~addition, as~most of the stars appear to host a super-Earth (or sub-Neptune), which is a type of planet that is not present in the Solar System, it seems that our planetary system architecture might not be much common. All the transit-based detection facilities in general and~TESS in particular show clear bias in the planetary systems found. Statistics from~\cite{nasa_exoplanet_archive} show that it is more common to find planetary systems with low orbital period, as~this produces a greater number 
of transits in the light curves. Additionally, planets with low planet-to-star radius ratio and semimajor axis are more~frequent.

Machine learning (ML) techniques allow for setting aside human judgment in order to detect and fit transit-like signals in light curves, thus reducing the overall time and computational cost required to analyze all of them. The~first approach was the Robovetter project~\cite{Coughlin_2016}, which was used to create \textit{the seventh Kepler planet candidate (PC) catalog}, employing a decision tree for classifying threshold crossing events (TCEs), which are periodic transit-like signals. Others, such as \textit{Signal Detection using Random forest Algorithm (SIDRA)} \cite{2016MNRAS.455..626M}, used random forests in order to classify TCEs depending on different features related to transit-like signals. Artificial neural networks (ANNs) were thought to be a better solution, in~particular convolutional neural networks (CNNs). The~results from~\cite{2018MNRAS.474..478P, Zucker_2018, Ansdell_2018, 2019MNRAS.488.5232C, Shallue_2018, gupta2023harnessing, Haider_2022, cuellar, Iglesias_2023} show that CNNs perform better in transit-like signal detection due to the fact that these algorithms are prepared for pattern recognition. One example was carried out in our previous work~\cite{Iglesias_2023}, in~which the performance of 1D CNN in transit detection was shown by training a model on K2 simulated data. Nowadays, as~current and future observing facilities provide very large datasets composed of many thousands of targets, automatic methods, such as CNN, are crucial for analyzing all of~them.

In this research, we continue to go a step further. As~our previous results showed, 1D CNNs are able to detect the presence of transit-like signals in light curves. The~aim of this research is to develop 1D CNN models that are able to extract different planetary parameters from light curves in which it is known that there are transit-like signals. There are other techniques that allow being automatized with ML techniques. One example was carried out in~\cite{Tasker_2020}, which studied how deep learning techniques can estimate the planetary mass from RV~data.

Our models were trained with simulated light curves that mimic those expected for the TESS mission, as~there are not enough confirmed planets detected with TESS (with their respective host star's light curves) to train a CNN model. The~reasons why simulated TESS light curves were used instead of K2 ones are motivated by the fact that apart from checking the performance of CNN in transit-like signal characterization, another aim is to check their performance with light curves from another survey. Our light curve simulator was adapted for creating both complete and phase-folded light curves. This comes from the fact that for phase-folding a light curve, first, it is necessary to know the orbital period of the planet (which is computed as the distance between two consecutive transits). After~knowing this parameter, a~phase-folded light curve could be computed by folding the complete light curve in the function of the orbital period. This is crucial because, taking into account the most common observing cadence from the main telescopes (about 30 min), a~transit with a duration of a few hours could be represented with only a few points, which makes the transit model fitting considerably difficult. By~phase-folding a light curve, the~transit shape is much more precise, as~it is composed of all the in-transit data points from the complete light curve. From~the complete light curves, our model is able to compute the orbital period, and~from the phase-folded light curve, it calculates the planet-to-star radius ratio and the semimajor axis in terms of its host star radius (see {Section~\ref{sec: train}}).

The rest of this paper is structured as follows: In {Section \ref{sec: maters}}, the materials and methods used during the research are explained. More concretely, in~{Section \ref{sec: shape}}, the~theoretical transit shape, which appears in all the light curves with which our models are trained, is explained; in {Section \ref{Sec: lc_sim}}, light curve simulation is detailed; and in {Section \ref{sec: models}}, the~structure of our model is shown in detail. In~{Section \ref{sec: train}}, the~training, test, and validation processes of the model are explained, and~also the statistics related to the predicted values obtained during the test process and the model test on real TESS data are shown and discussed. Finally, in~{Section \ref{sec: conclusions}}, the~conclusions of all the research are~outlined.

\section{Materials and~Methods}\label{sec: maters}
In this section, the~main materials and methods used during this research are introduced, including the explanation of the transit theoretical shape, the~light curve simulation, and CNNs, placing special emphasis on the models~used.

\subsection{Transit~Shape}\label{sec: shape}

\begin{figure}[H]
 \hspace{-0.5cm}   \includegraphics[width=0.98\linewidth]{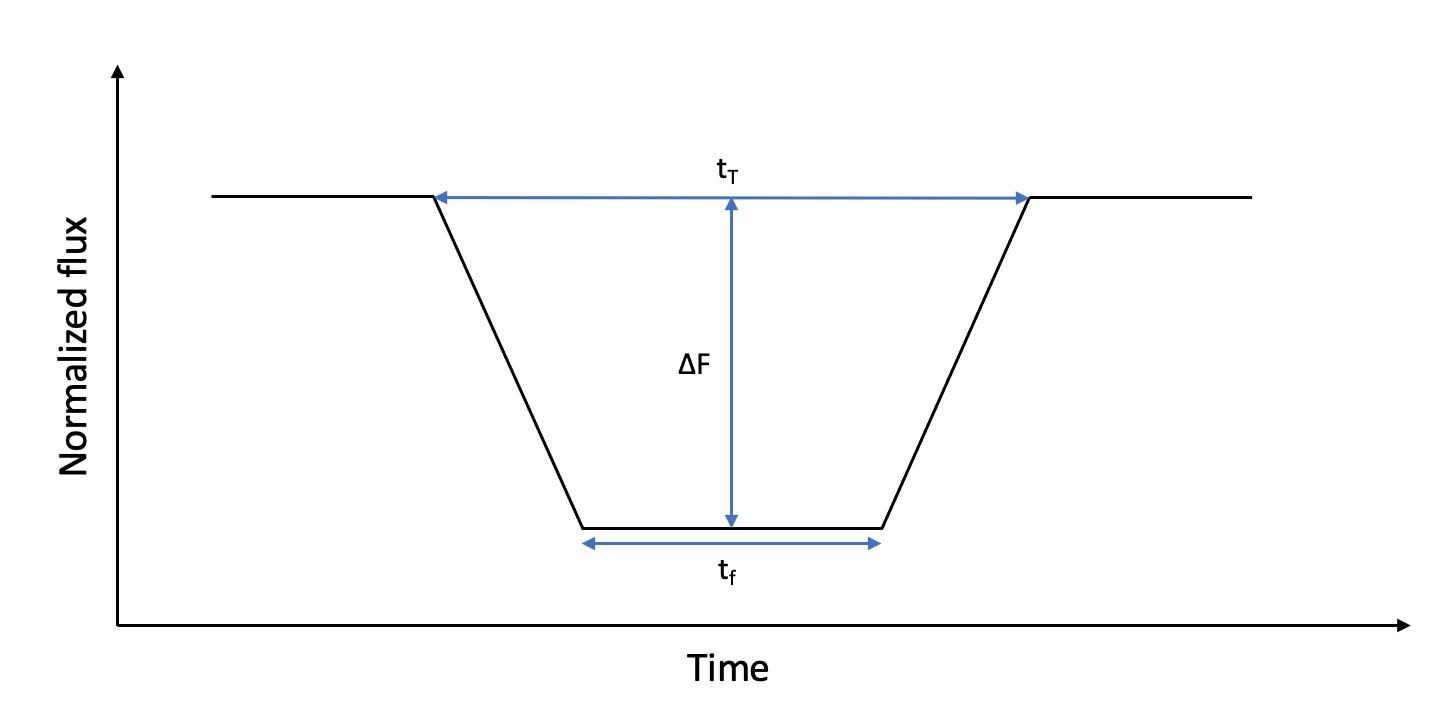}
    \caption{Theoretical transit model without limb~darkening.}
    \label{fig: theor_model}
\end{figure}

Theoretically, a~transit can be described as a trapezoid where the amount of flux that reaches the detector decreases while the planet overlaps the star. There are 3 main parameters that can be derived from its shape: the transit depth ($\Delta F$), the~transit duration ($t_T$), and~the duration of the flat region ($t_F$), which is observed when the whole planet is overlapping the star (see {Figure~\ref{fig: theor_model}}).

\vspace{-9pt}

The flux that arrives at the detector, which is monitoring the star, could be expressed as~\cite{Mandel_2002}
\begin{equation}
F(a[r_\star], R_p[R_\star])=1-\lambda(a[r_\star], R_p[R_\star])
\label{eqn: shape_1},
\end{equation}
where $F(a[r_\star], R_p[R_\star])$ is the flux that reaches the telescope and $\lambda(a[r_\star], R_p[R_\star])$ is the portion of the flux that is blocked by the planet. As~the planet overlaps the star, the~amount of flux~decreases.

Actually, the~transit shape is more complex as there is an optical effect that makes the star appear less bright at the edges than at the center, known as limb darkening. This effect causes the transit to be rounded. The~most accurate transit shape, taking into account this phenomenon, is described in~\cite{Mandel_2002} (Mandel and Agol theoretical shape). The~limb darkening formulation adopted in the Mandel and Agol shape is described in~\cite{2000A&A...363.1081C}, where how the intensity emitted by the star changes in the function of the place where the radiation comes from is described:
\begin{equation}
\frac{I(\mu)}{I(0)}=1-\sum^4_{m=1} c_m (1-\mu^{m/2})
\label{eqn: claret},
\end{equation}
with $\mu=\sqrt{1-\tilde{r}^2}$, where $\tilde{r} \in [0,1]$ is the normalized radial coordinate on the disk of the star, $c_m$ are the limb darkening coefficients, and \textit{I}(0) is the intensity at the center of the star. Another common parameterization of this phenomenon is made with a quadratic law~\cite{1950HarCi.454....1K}, where $u_1$ and $u_2$ are the quadratic limb darkening coefficients:
\begin{equation}
\frac{I(\mu)}{I(0)}=1-u_1\cdot(1-\mu)-u_2\cdot(1-\mu)^2.    
\end{equation}

\begin{figure}[H]
    \includegraphics[width=0.98\linewidth]{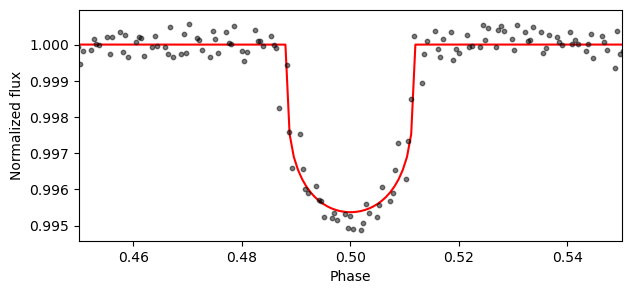}
    \caption{Example of a transit shape with the Mandel and Agol model computed with TLS from the simulated light curve shown in the upper panel of {Figure~\ref{fig: lcs}}. Black: phase-folded light curve. Red: TLS~model.}
    \label{fig: mandel_shape}
\end{figure}

Taking into account the limb darkening effect, the~Mandel and Agol theoretical shape models transits as
\begin{equation}
F(a[r_\star], R_p[R_\star])=\left[ \int^1_0 dr 2 r I(r)\right]^{-1} \int^1_0drI(r)\frac{d[\tilde{F}(a[r_\star]/r, R_p[R_\star]/r])}{dr}
\label{eqn: shape_2},
\end{equation}
where $\tilde{F}(a[r_\star]/r, R_p[R_\star]/r])$ is the transit shape without the limb darkening effect. An~example of the Mandel and Agol transit shape obtained with TLS from a simulated light curve (see {Section \ref{Sec: lc_sim}}) is shown in Figure~\ref{fig: mandel_shape}. The~value $t_f$ is much more difficult to determine. However, as~was previously mentioned, it is still possible to estimate different parameters, as~in the case of $ R_p[R_\star]$, which is related to the squared root of the mean depth of the transit. For~this reason, it is thought that convolutional neural networks, which learn from the input dataset shape, could be a solution. They could infer the main parameters from a phase-folded light curve. Others, as~the orbital period, should be inferred from the complete light curve, because~its value is the mean distance between two consecutive~transits.

\unskip

\subsection{Light Curve~Simulation}\label{Sec: lc_sim}
The aim of this research is the creation of a CNN architecture that is able to predict the values $P$, $R_p[R_\star]$, and $a[R_\star]$ and that is trained with TESS light curves. However, as~a CNN needs a large dataset for training all its hyperparameters, and~there are not enough real light curves with confirmed planetary transits to train it, it was decided to use simulated light curves to train and test the models. This is why the light curve simulator used in our previous work was adapted to TESS data. It is important to remark that TESS light curves are shorter than K2 ones due to the fact that TESS observes a sector during 27~days instead of 75 days (which is the mean duration of K2 campaigns). In~contrast, as~both of them have a mean observing cadence of 30 min, the~observing cadence was not changed. However, less bright stars were selected in order to increase the noise levels, which would mean that the network would also be able to learn to characterize transits where noise levels make this work considerably much more~difficult.

Before explaining our light curve simulator, it is necessary to clarify that stellar light curves are affected by stellar variability phenomena, such as rotations, flares, and pulsations, which induce trends that have to be removed before analyzing them. The~light curves after removing these trends are usually known as detrended (or normalized) light~curves.

For simulating light curves, the~Batman package~\cite{Kreidberg_2015}, which allows for creating theoretical normalized light curves with transits (i.e., detrended light curves with transits but without noise), was used. The~main parameters needed for creating the transit models are the orbital period (P); the~planet-to-star radius ratio ($R_p[R_\star]$); the~semimajor axis in terms of the host star's radius ($a[r_\star]$); the~epoch ($t_0$), which is the time in which the first transit takes place; the~inclination angle of the orbital plane ($i[deg]$); and the limb darkening quadratic coefficients ($u_ 1$ and $u_2$).

All these parameters were selected, when it was possible, as~random values, where the upper and lower limits were chosen, taking into account planetary statistics (in general, not only the TESS ones) or the main characteristics of TESS light curves. As~previously mentioned, as~the TESS mission detects exoplanets through the transit method, it is more common to find planetary systems with low periods, planet-to-star radius ratio, and semimajor axis. However, as~the main objective is to develop a model that is 
as most generalized as possible, it was decided not to use these statistics, thus preventing the~model from generating dependence and not being able to correctly characterize systems with different parameters. 
First of all, it was decided to inject at least 2 transits in the light curves (necessary to be able to estimate the orbital period from light curves), so the maximum value of the period considered is half of the duration of the light curves. The~value is chosen randomly between these limits. This is the only restriction applied that generates bias in the data, but~it is necessary to apply it because if there were not at least 2 transits in the light curves, the~model would not be able to estimate the orbital period. Then, the~epoch was chosen as a random value between 0 and the value of the orbital~period.

For choosing $R_p[R_\star]$, the following procedure was carried out: First of all, it is necessary to simulate the host star. The~stellar mass was chosen as a random value following the statistics of the solar neighborhood, which is the surrounding region to the Sun within $\sim$92 pc. Their radii were estimated following the mass--radius relationship of the main sequence (MS) stars~\cite{demory, 1991Ap&SS.181..313D} (among others):
\begin{equation}
    R_\star[R_\odot]\sim (M_\star[M_\odot])^{0.8}
\end{equation}

The distance and apparent magnitude (which is a measure of the brightness of a star observed from the Earth on an inverse logarithmic scale) were chosen as random values limited, respectively, by~the solar neighborhood size (up to $\sim$92 pc) and TESS typically detected magnitudes (up to magnitude 16) \cite{2015JATIS...1a4003R}. Limb darkening quadratic parameters ($u_1$ and $u_2$) are chosen as random values between 0 and 1 considering $u_1+u_2=1$. Then, the~planetary radius was chosen following planetary statistics: it is not common to find Jupiter-like exoplanets orbiting low mass stars as red dwarfs, so the maximum radius of the orbiting exoplanet was limited to the one expected from a Neptune-like one in low mass stars (i.e., the~maximum $R_p[R_\star]$ was set to $0.05$ for stars with masses lower than $0.75~M_\odot$). For~more massive stars, the~upper limit is a Jupiter-like exoplanet (i.e., the~maximum value is limited to $R_p[R_\star]= 0.1$ if the stellar mass is larger than $0.75~M_\odot$). $a[r_\star]$ was derived considering Kepler's third law:
\begin{equation}
    a=\left(\frac{G\cdot M_\star}{4\cdot \pi^2}\cdot P^2\right)^\frac{1}{3}
\label{eqn: Kepler}.
\end{equation}

To summarize all the stellar and planetary parameters used during the light curve simulation, the~upper and lower limits from all of them are shown in {Table~\ref{tab: priors}}.

\begin{table}[H]
\centering
\caption{Stellar and planetary paramaters' upper and lower limits used for simulating the light~curves.}
\begin{tabularx}{\textwidth}{CCC}
\toprule
\textbf{Parameter}             & \textbf{Min Value} & \textbf{Max Value} \\ \midrule
$M_\star[M_\odot]$ & 0.1       & 10        \\ \midrule
$R_\star[R_\odot]$  & 0.15      & 4.89      \\ \midrule
$u_1$              & 0         & 1         \\ \midrule
$u_2$              & 0         & 1         \\ \midrule
mag               & 8         & 16        \\ \midrule
$d[pc]$           & 1         & 92       \\ \midrule
$P(d)$            & 1         & 13.5      \\ \hline
$t_0$             & 0         & $P(d)$    \\ \hline
$R_p[R_\star]$    & 0.01      & 0.1      \\ \bottomrule
\end{tabularx}
\label{tab: priors}
\end{table}

As was already mentioned, phase-folded light curves, which allowed for obtaining more accurate transit shapes, thus producing better performance in transit characterization, were also simulated. The~main difference when creating both types of light curves is that in the complete light curve, the~temporal vector needed for simulating them takes into account the whole duration of a TESS sector; on~the contrary, a~phase-folded light curve is created with a temporal vector between $t_0-P/2$ and $t_0+P/2$, which other authors call as a \textit{global} view of the transit (a local version will be created by making a zoom to the transit) \cite{Shallue_2018, Ansdell_2018} (among others).

Batman creates light curves without noise, so it has to be added after generating them. For~that aim, Gaussian noise was implemented following the expected values for TESS light curves; i.e.,~the standard deviation ($\sigma$) expected from TESS light curves in the function of the stellar magnitude~\cite{2015JATIS...1a4003R} was taken into account, and thus, a~\textit{noise vector}, which is centered in 0, was obtained. As~the light curve without noise has a maximum of {1}, the~light curve with noise was created by adding both vectors. The~signal-to-noise ratio (SNR) for each of the detrended light curves in the function of its stellar magnitude was estimated with the relationship shown in {Equation~\eqref{eqn: snr}} \cite{schroeder1999astronomical}, where $\mu=1$ is the mean value (after adding both vectors) and $\sigma$ is the standard deviation proportional to stellar magnitude~\cite{2015JATIS...1a4003R}. In~addition, key SNR values in the function of stellar magnitude are shown in {Table~\ref{tab: SNR}}. {It is important to remark that the~larger the magnitude is, the~lower its SNR will be, and thus, the~transit-like signal detection and characterization will also be more difficult.}
\begin{equation}
    SNR=10\cdot log_{10}(\frac{\mu}{\sigma})
    \label{eqn: snr}
\end{equation}

\begin{table}[H]
\centering
\caption{SNR in the function of stellar~magnitude.}
\begin{tabularx}{\textwidth}{CC}
\toprule
\textbf{mag} & \textbf{SNR}     \\ \midrule
8   & 40.84 \\ \midrule
10  & 36.20 \\ \midrule
12  & 31.56 \\ \midrule
14  & 26.92  \\ \midrule
16  & 22.28 \\ \bottomrule
\end{tabularx}
\label{tab: SNR}
\end{table}

\begin{figure}[H]
   \includegraphics[width=0.98\columnwidth]{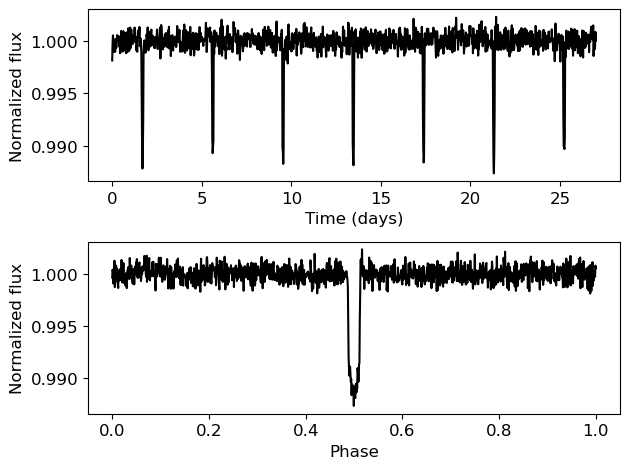}
    \caption{Example of the 2 views of a simulated light curve. \textbf{Top panel}: complete light curve. \textbf{Bottom panel}: phase-folded light curve. Phase is computed as $(t-t_0)/P$.}
    \label{fig: lcs}
\end{figure}

\begin{figure}[H]
  \includegraphics[width=\columnwidth]{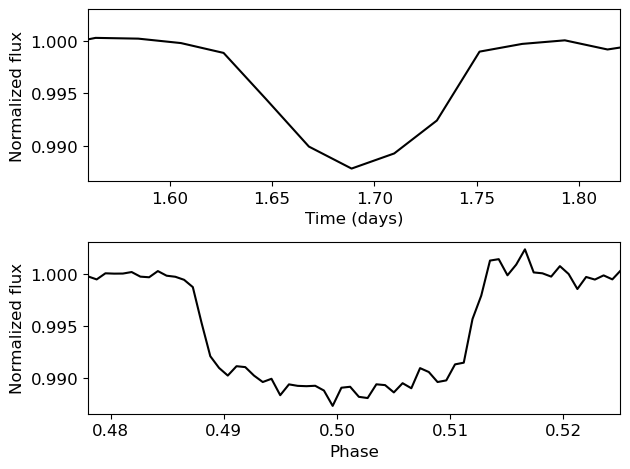}
    \caption{Zoom of the light curves of Figure~\ref{fig: lcs}. \textbf{Top panel}: complete light curve. \textbf{Bottom panel}: phase-folded light~curve.}
    \label{fig: lcs_zoom}
\end{figure}

\begin{figure}[H]
  \includegraphics[width=\columnwidth]{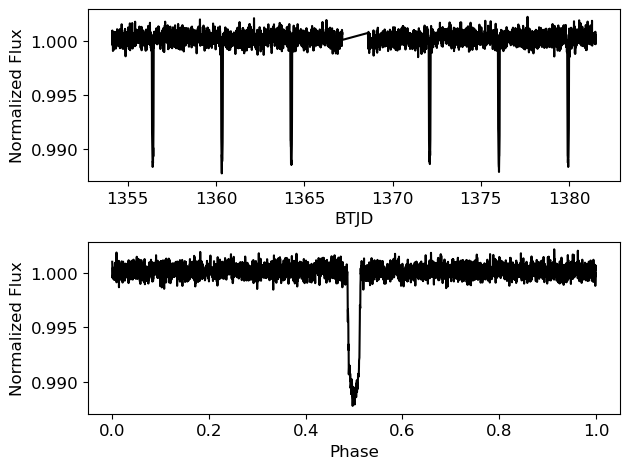}
    \caption{TOI-192 light curve. \textbf{Top panel}: Full light curve. \textbf{Bottom panel}: phase-folded light curve computed with TLS.}
    \label{fig: real}
\end{figure}

An example of the 2 views (complete and phase-folded) of a simulated light curve is shown in {Figure~\ref{fig: lcs}}. As~shown, a~phase-folded light curve allows for obtaining a better transit shape. For~checking this, a~zoom of {Figure~\ref{fig: lcs}} is shown in {Figure~\ref{fig: lcs_zoom}}. As~shown, the~transit of the phase-folded light curve is composed of much more points, thus allowing for better checking the Mandel and Agol theoretical shape. Both views of the light curve were simulated, taking into account the main parameters from TIC 183537452 (TOI-192), except~the epoch and the temporal span of the light curve, which were not considered as they do not condition the noise levels and the transit shape. TOI-192 is part of the TOI Catalog, short for the TESS Objects of Interest Catalog, which comprises a compilation of the most favorable targets for detecting exoplanets. Its complete and phase-folded light curves are shown in {Figure~\ref{fig: real}}. TESS light curves exhibit gaps that occur during the data download by the telescope. However, this phenomenon was not considered because either no transits are lost in the gap, and thus, the~separation between two consecutive transits is maintained, or if~any transits are in this zone, the~distance between the last transit before the gap and the next one after the gap will be separated by a temporal distance proportional to the orbital period, so the calculation of this value will not be affected. Aside from this, the~simulated light curves closely resemble those expected from TESS, showing similar noise levels for similar stellar magnitudes and showing a similar transit depth similar to $(R_p[R_\star])^2$.

The training process (see {Section \ref{sec: train}}) requires 2 different datasets, as~it is important to test both models in order to check their generalization to data unknown for the CNN (this is known as test process). Thus, a~train dataset with 650,000 light curves and a test dataset with 150,000 were developed. The~number of light curves in the train dataset was chosen after training both models a large number of times and increasing the number of light curves used. The~final amount of light curves corresponds to the one with which the best results were obtained. Above~this number, significant improvement was seen. In~addition, the~number of light curves in the test dataset was chosen to have a large statistic that would allow a good check of the~results.

\unskip

\subsection{Convolutional Neural Networks (CNNs): Our 1D CNN Models}\label{sec: models}

The transit shape previously shown differs from an outlier or the noise present in light curves only by their shape. This is why CNNs play a crucial role in transit detection and characterization, as~these algorithms could distinguish between both signals (noise and transits), even though they are at similar~levels.

\begin{figure}[H]
\hspace{-0.8cm}\includegraphics[width=0.8\linewidth]{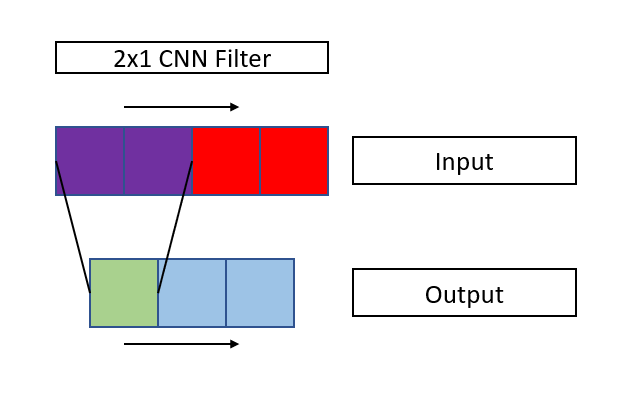}
\caption{Example of how a 1D CNN filter works. Red: input. Blue: output. Purple: filter input. Green: filter output (after applying the appropriate operation). Black arrows show the movement of the~filter.}
\label{fig: 1D_filter}
\end{figure}

CNNs apply filters to the input data that allow the detection of different features that characterize them. An~example of a 1D CNN filter is shown in {Figure~\ref{fig: 1D_filter}}.

\vspace{-9pt}

\textls[-5]{The main activation function used in this research is known as parametric rectifier linear unit (PReLU) \cite{2015arXiv150201852H}}. If $y_i$ is considered as the input of the non-linear activation function, its mathematical definition is:
\begin{equation}
    f(y_i)=\left\{ 
    \begin{array}{lr}
    y_i & if \:\: y_i>0 \\
    a_i\cdot y_i & if \:\: y_i\leq 0
    \end{array}
    \right.
\label{eqn: PReLU}.
\end{equation}

\textls[-15]{The main difference between this activation function and the rectifier linear unit (ReLU)~\cite{2018arXiv180308375A}}, which is one of the most commonly used, is that, in~ReLU, $a_i$ is equal to 0, while in PReLU, it is a learnable~parameter.

The gradient of this function when optimizing with backpropagation~\cite{6795724} is (taking into account a derivative with respect to the new trainable parameter, $a_i$)
\begin{equation}
    \frac{\partial f(y_i)}{\partial a_i}= \left\{ 
    \begin{array}{lr}
    0 & if \:\: y_i>0 \\
    y_i & if \:\: y_i\leq 0
    \end{array}
    \right.
\label{eqn: grad_PReLU}.
\end{equation}

Our CNN model consists, actually, in~two 1D CNN models, as~a transit analysis could be understood as a temporal series one. The~main fact is that flux vectors of the simulated light curves were used  as inputs for training, validating, and testing them (see {Section \ref{sec: train}}). All the processes were carried out in Keras~\cite{chollet2015keras}. The~first model (it will be referred to as model 1 from~now on) works with complete light curves and is able to predict the value of the orbital period of the transit-like signals. The~second model (it will be referred to as model 2 from~now on) works with phase-folded light curves and is able to predict $(R_p[R_\star])^2$ and $a[R_\star]$. The~model predicts $(R_p[R_\star])^2$ instead of $R_p[R_\star]$ because the squared value is related to the transit~depth.

\begin{figure}[H]
   \includegraphics[width=0.65\linewidth]{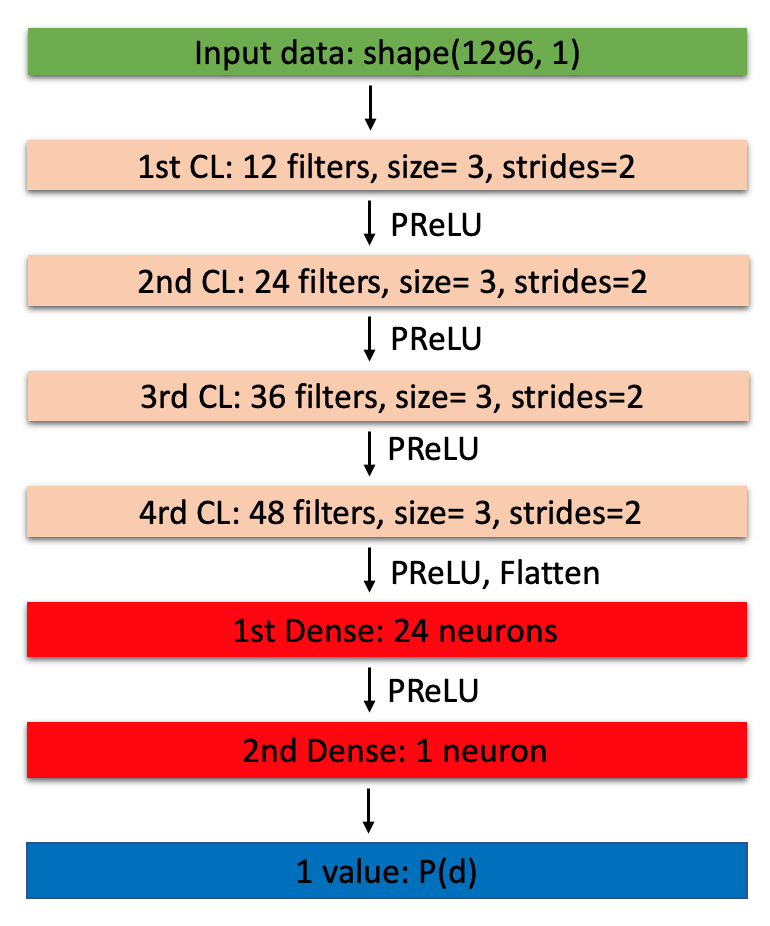}
    \caption{Scheme of model 1. Green: input. Salmon: convolutional layers. Red: MLP layers. \mbox{Blue: output.}}
    \label{fig: modelo_1}
\end{figure}

The structure of both models is the same with the difference of the last layer, which has the same number of neurons as the number of parameters to predict. Both models are composed of a 4-layer convolutional part and a 2-layer multilayer perceptron (MLP) part. In~the convolutional part, all the layers have a filter size of 3 and have 2 strides. The~numbers of convolutional filters are, respectively, 12, 24, 36, and 48. All the layers are connected by a PReLU activation function and have the padding set to the \textit{same}, which avoids the change of the shape of the light curves among the layers due to the convolutional filters. The~MLP part is composed of a layer with 24 neurons activated by PReLU and a last layer with 1~neuron for model 1 and 2 neurons for model 2, without~an activation function. The~two parts of the models are connected by a flatten layer. A~scheme of model 1 is shown in {Figure~\ref{fig: modelo_1}}. The~scheme of the second model consists of changing the number of neurons to 2 in the last layer (2nd sense). \\

\vspace{-9pt}
\unskip

\section{Model Training and Test 
Results and~Discussion}\label{sec: train}
First of all, both processes were carried out on a server with an Intel Xeon E4-1650 V3, 3.50 GHz, with~12 CPUs (6 physical and 6 virtual). It had a RAM of 62.8~Gb.

As both models are 1D CNN models, it is important to preprocess the input light curves due to the fact that convolutional filters usually take the maximum value on them, which would entail the loss of the transits. Thus, the~light curves were inverted and set between 0 and~1. 

In addition, the~values related to the parameters that the 1D CNN models were learning to predict (from now on, it will be referred as labels) were normalized. The~following transformation was applied to each label, supposing \textit{max\_labels} as the maximum of all the labels simulated and \textit{min\_labels} the minimum value:
\begin{equation}
    label=\frac{label-min\_labels}{max\_labels-min\_labels}.
\end{equation}

These transformations set the data between 0 and 1. This is necessary because deep learning techniques work more properly when the input and output data are~normalized.

Before training both models, adaptive moments (Adam) \cite{2014arXiv1412.6980K} were selected as the optimizer. Adam is a stochastic gradient  descent (SGD) \cite{10.1214/aoms/1177729586} method that changes the value of the learning rate using the values of the first- and second-order gradients. The~initial learning rate was set to 0.0001 (chosen after many training processes). As~a loss function, the~mean squared error (MSE) was~chosen. 

One of the main parts of the training process consists in validating the model train with a dataset different from the one used for updating the hyperparameters in each epoch (an epoch is each time the training dataset is used to train the model) in order to check how well the model is being generalized during the training process. Keras allows for performing that with a parameter known as \textit{validation split}. A~number between 0 and 1 that refers to the percentage of the train dataset is used for this process. A proportion of 30\% of the dataset was split; i.e.,~a \textit{validation split} of 0.3 was selected. Furthermore, a~batch size of 16 was selected. For~monitoring the training, the MSE loss function was chosen. Each epoch took about 200 s to be completed in our~server.

\begin{figure}[H]
    \includegraphics[width=0.8\linewidth]{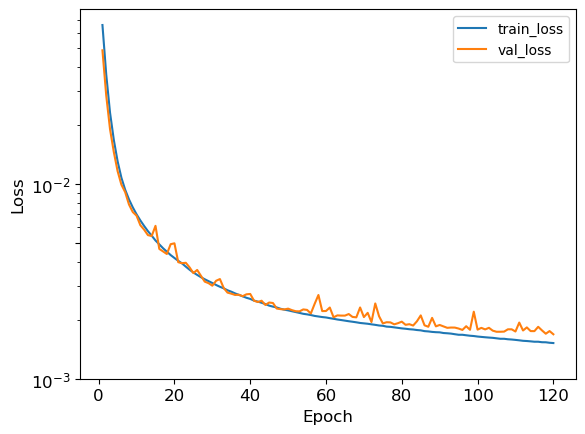}
    \caption{Training history of model 1. Blue: training loss. Orange: validation~loss.}
    \label{fig: hist_1}
\end{figure}

\begin{figure}[H]
   \includegraphics[width=0.8\linewidth]{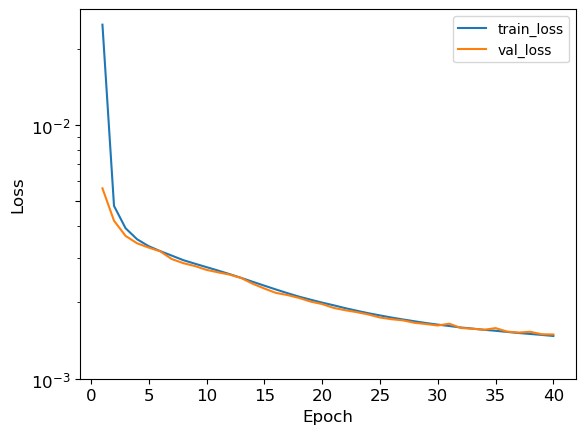}
    \caption{Training history of model 2. Blue: training loss. Orange: validation~loss.}
    \label{fig: hist_2}
\end{figure}

After training both models, the~training histories were obtained, in~which training and validation loss were plotted against the epochs. {Figures \ref{fig: hist_1}} and {\ref{fig: hist_2}} show, respectively, the~training history of models 1 and 2 in a logarithmic scale. As~shown, the~training process was completed correctly, as~the validation loss decreased along with the training loss among the~epochs.

\begin{figure}[H]
   \includegraphics[width=0.95\linewidth]{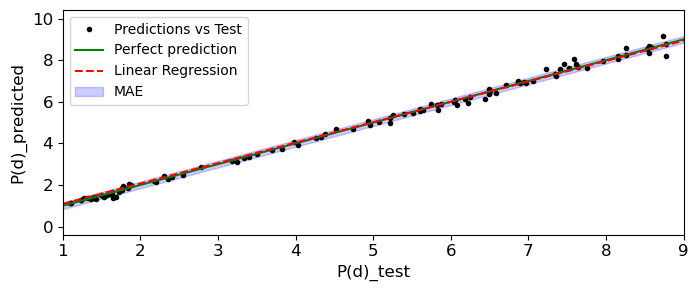}
    \caption{Comparison between 100 predicted and test values (black dots) from a sample of $P$ of 100. Red: linear regression. Dashed blue area: MAE. Green: perfect prediction. Shaded blue: MAE.}
    \label{fig: plot_1}
\end{figure}

\begin{figure}[H]
    \includegraphics[width=\linewidth]{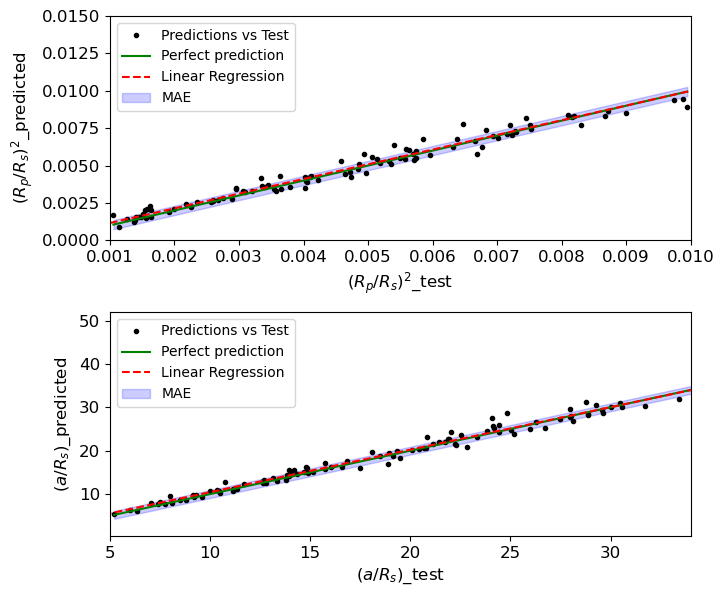}
    \caption{Comparison between 100 predicted and test values (black dots) from a sample of $(R_p[R_\star])^2$ (\textbf{upper panel}) and $a[R_\star]$ (\textbf{bottom panel}) of 100. 
    Red: linear regression. Dashed blue area: MAE. Green: perfect prediction. Shaded blue: MAE.}
    \label{fig: plot_2}
\end{figure}

\begin{figure}[H]
\includegraphics[width=\linewidth]{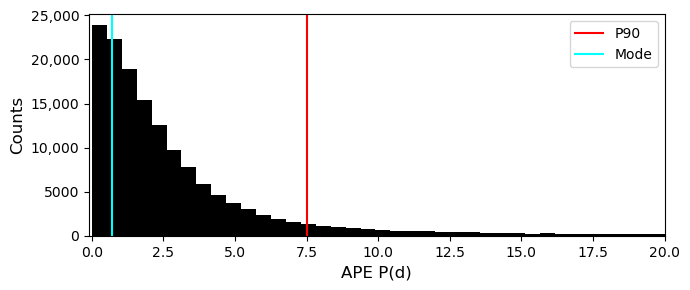}
 \caption{APE histograms for $P(d)$. Cyan: mode. Red: percentile 90 (P90).}
\label{fig: histo_1}
\end{figure}

\begin{figure}[H]
 \includegraphics[width=\linewidth]{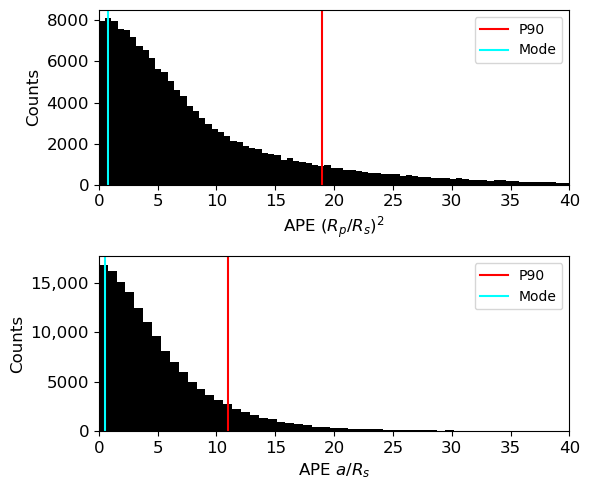}
 \caption{APE histograms for $(R_p[R_\star])^2$ (upper panel) and $a[R_\star]$ (lower panel). Cyan: mode. Red: percentile 90 (P90).}
 \label{fig: histo_2}
\end{figure}

The test process was carried out with the test dataset composed by 150,000 light curves (as explained in {Section \ref{Sec: lc_sim}}). The~accuracy of the predictions was studied with a comparison between them and the values with which the light curves were simulated. Theoretically, the~perfect result would be adjusted to a linear function with a slope 1 and an intercept of 0. The~results were fit with a linear regression and obtained $R^2= 0.991$ for the orbital period (model 1), $R^2=0.974$ for $(R_p[R_\star])^2$ (model 2), and $R^2=0.971$ for $a[R_\star]$ (model 2). These results mean that most of the data predicted are in agreement with the simulated data, which implies that our 1D CNN models are properly predicting the planetary parameters. Furthermore, the~mean absolute error (MAE) was computed for both models, obtaining $MAE= 0.015d$ for the orbital period (model 1), $MAE= 0.0003$ for $(R_p[R_\star])^2$ (model 2), and $MAE= 0.9113$ for $a[R_\star]$ (model 2). These results could be better understood with the plot of a small randomly chosen sample of 100 predictions and test data (see \mbox{{Figures \ref{fig: plot_1}} and { \ref{fig: plot_2}}}) because the total plot of the data is a bit clumpy and could make the results confusing. As~shown, the~predictions fit correctly to the simulated ones. In~addition, as~an absolute error could be a bit confusing, the~absolute percentage error (APE) was computed from all the predictions and the obtained value were plotted in histograms, in~this case, with the whole data (see \mbox{{Figures \ref{fig: histo_1}} and {\ref{fig: histo_2}}}, which make reference, respectively, to~models 1 and 2). The~values of the modes from the histograms of $P(d)$, $(R_p[R_\star])^2$ and $a[R_\star]$ are, respectively, 0.003, 0.69, and 0.38. To~sum up all the results, all of them are shown in {Table~\ref{tab: tab_1}}.

\vspace{-9pt}

\begin{table}[H]
\caption{Statistics computed with the predictions of the test process for each of the predicted parameters (P(d), $(R_p[R_\star])^2$, and $a[R_\star]$).}

\begin{tabularx}{\textwidth}{CCCcc}
\toprule
\textbf{Parameter}          & \textbf{MAE}    & \boldmath{$R^2$} &  \textbf{Histogram Mode (\%) }& \textbf{Histogram P90 (\%)} \\ \midrule
P(d)               & 0.015   & 0.99 & 0.7                 & 7.5                \\ \midrule
$(R_p[R_\star])^2$ & 0.0003 & 0.97 & 0.8                 & 19                 \\ \midrule
$a[R_\star]$       & 0.91 & 0.97 & 0.5                 & 11                 \\ \bottomrule
\end{tabularx}
\label{tab: tab_1}
\end{table}

\vspace{-9pt}

\unskip

\unskip

All these results show that both models properly predict the parameters \textit{P}, $(R_p[R_\star])^2$ and $a[R_\star]$, with~low uncertainties. In~addition, they show that both models properly generalize without generating dependence on the train dataset. Among~other previous studies, these results show that 1D CNN is a good choice not only for checking if a light curve presents transit-like signals but also for estimating parameters related to its shape. The~test dataset shows a wide range of values, but~even so, the~models do not show any bias in the results (which would mean that the models predict better in some regions than in others), which means that the models are properly trained. To~put these results in context, 10,000 light curves from the test dataset were analyzed with TLS and the $MAE$ for $P(d)$ and $(R_p[R_\star])^2$ were computed respect to the values with which the light curves were simulated (TLS does not compute the semimajor axis of the orbit). The~obtained values are, respectively, $0.009$ and $0.0018$, which mean that our CNN models predict all the parameters with a similar precision compared with the most used algorithms~nowadays.

\begin{figure}[H]
    \includegraphics[width=\linewidth]{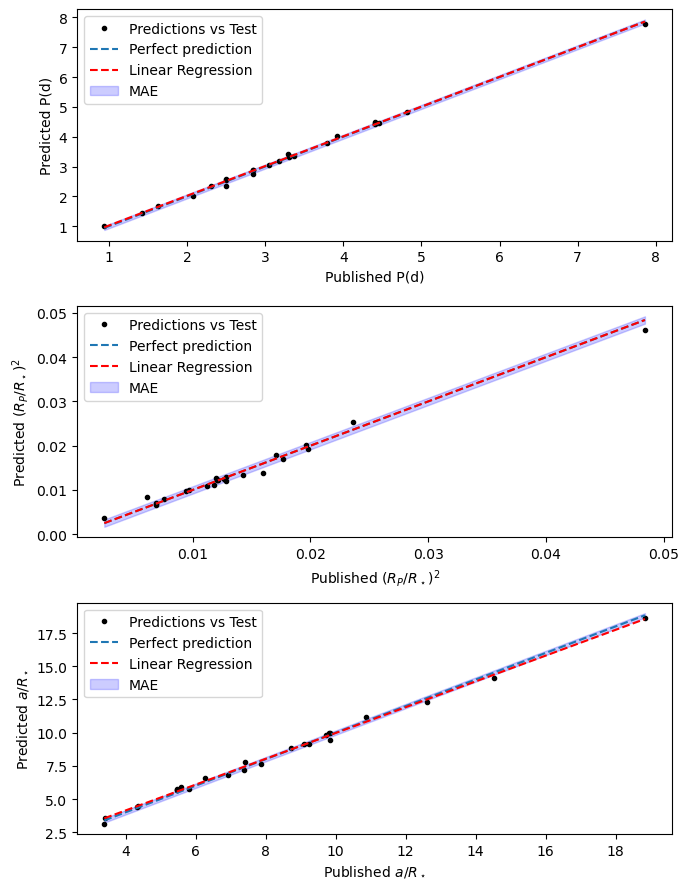}
    \caption{Predicted parameters on real TESS data. Prediction vs. real data (black dots) of $P(d)$ (\textbf{upper panel}), $(R_p[R_\star])^2$ (\textbf{middle panel}), and $a[R_\star]$ (\textbf{bottom panel}). Red: linear regression. Dashed blue area: MAE. Green: perfect prediction. Shaded blue: MAE.}
    \label{fig: real_plots}
\end{figure}

The models were also tested on real TESS data. From~the Mikulski Archive for Space Telescopes (MAST) (bulk download \url{https://archive.stsci.edu/tess/bulk_downloads/bulk_downloads_ffi-tp-lc-dv.html} (accessed on December 2023)), 25 light curves with confirmed transiting exoplanets from the stars part of the \textit{candidate target lists (CTLs)} \cite{Stassun_2018}, which is a special subset from the \textit{TESS Input Catalog (TIC)} containing targets that are good for detecting planetary-induced transit-like signals, were obtained. This subset of light curves was decided to be used because there are preprocessed and corrected light curves available to download from all of them. All the light curves were analyzed with our \textit{1D CNN} models, and the results were compared with their published values from~\cite{Tasker_2020}. The~phase-folded light curves were computed with the predicted value of the orbital period. From~the results, the~\textit{MAE}, $R^2$, and \textit{Mean Absolute Percentage Error (MAPE)} were computed. All the predicted and real values are shown in {Table~\ref{tab: preds}}, along with the APE and the Absolute Error (AE) computed for each of the parameters. The~obtained \textit{MAEs} of \textit{P(d)}, $(R_p[R_\star])^2$, and $a[R_\star]$ are, respectively, $0.0502 $, $0.0004$, and $0.20$; the \textit{MAPE} of \textit{P(d)}, $(R_p[R_\star])^2$, and $a[R_\star]$ are, respectively, $1.87\%$, $7.16\%$, and $2.98\%$; and the $R^2$ values from $P(d)$, $(R_p[R_\star])^2$, and $a[R_\star]$ are, respectively, $0.981$, $0.978$, and $0.985$. These results mean that our \textit{1D CNN} models also perform properly on real data. In~{Figure~\ref{fig: real_plots}}, the~comparison between real data and predicted values is plotted to check the accuracy of the predictions visually. The~differences between the results obtained with real and simulated data were due to the fact that, although~simulating the light curves as most similar as possible to those expected for the TESS mission was attempted, in~reality, it is impossible to have them the same. However, these results show that both models are able to characterize transiting exoplanets from the TESS light curves of their host star also in real data, which means that our models were well trained and that the light curve simulator is able to mimic real TESS light curves with high~accuracy.

\startlandscape
\begin{table}[H] 
\caption{Published and predicted parameters of exoplanets observed by TESS. The~first two columns show, respectively, the~TIC id of the star and the sector in which each light curve was obtained. 
The~other columns show the published parameters, the~predicted parameters, the~AE, and the APE of each of the parameters and exoplanets. Published data were obtained from~ExoFOP.}
\tablesize{\footnotesize}
\begin{tabularx}{\textwidth}{ccccccCCcccccc}
\toprule
\textbf{TIC id}    & \textbf{Sector} & \textbf{P(d) Real} & \textbf{P(d) Pred} & \textbf{AE (P(d))} & \textbf{APE (P(d))(\%)} & \boldmath\textbf{$(R_P[R_\star])^2$ Real} & \boldmath\textbf{$(R_P[R_\star])^2$ Pred} & \boldmath\textbf{AE ($((R_P[R_\star])^2))$} & \boldmath\textbf{APE ($((R_P[R_\star])^2)$) (\%)} & \boldmath\textbf{$a[R_\star]$ real} & \boldmath\textbf{$a[R_\star]$ Pred} & \boldmath\textbf{AE $(a[R_\star])$} & \boldmath\textbf{APE $(a[R_\star]) (\%)$} \\ \midrule
183537452 & 2      & 3.9230    & 4.0200    & 0.0970   & 2.4726    & 0.0094                  & 0.0098                  & 4.2553                  & 4.2553                   & 12.6200           & 12.3000           & 0.3200           & 2.5357                                                       \\ \midrule
158623531 & 2      & 7.8600    & 7.7900    & 0.0700   & 0.8906    & 0.0127                  & 0.0125                  & 1.2598                  & 1.2598                   & 14.5200           & 14.1000           & 0.4200           & 2.8926                                                       \\ \midrule
100100827 & 2      & 0.9414    & 1.0150    & 0.0736   & 7.8181    & 0.0097                  & 0.0100                  & 3.0928                  & 3.0928                   & 3.3900            & 3.1500            & 0.2400           & 7.0796                                                       \\ \midrule
388104525 & 2      & 2.4997    & 2.5762    & 0.0765   & 3.0604    & 0.0128                  & 0.0130                  & 1.5625                  & 1.5625                   & 7.4200            & 7.7900            & 0.3700           & 4.9865                                                       \\ \midrule
149603524 & 2      & 4.4119    & 4.4986    & 0.0867   & 1.9651    & 0.0121                  & 0.0122                  & 1.0744                  & 1.0744                   & 5.5700            & 5.9000            & 0.3300           & 5.9246                                                       \\ \midrule
184240683 & 2      & 1.6284    & 1.6700    & 0.0416   & 2.5547    & 0.0122                  & 0.0123                  & 0.6987                  & 0.6987                   & 3.3970            & 3.5800            & 0.1830           & 5.3871                                                       \\ \midrule
38846515  & 2      & 2.8494    & 2.7500    & 0.0994   & 3.4885    & 0.0069                  & 0.0071                  & 3.0316                  & 3.0316                   & 5.4800            & 5.6600            & 0.1800           & 3.2847                                                       \\ \midrule
230982885 & 2      & 2.0728    & 2.0000    & 0.0728   & 3.5122    & 0.0112                  & 0.0108                  & 3.5714                  & 3.5714                   & 9.7300            & 9.8500            & 0.1200           & 1.2333                                                       \\ \midrule
149603524 & 1      & 4.4119    & 4.4600    & 0.0481   & 1.0902    & 0.0121                  & 0.0121                  & 0.3306                  & 0.3306                   & 6.2800            & 6.5600            & 0.2800           & 4.4586                                                       \\ \midrule
38846515  & 1      & 2.8494    & 2.8800    & 0.0306   & 1.0739    & 0.0069                  & 0.0065                  & 5.7971                  & 5.7971                   & 9.8376            & 9.9961            & 0.1585           & 1.6113                                                       \\ \midrule
388104525 & 3      & 2.4997    & 2.3400    & 0.1597   & 6.3888    & 0.0118                  & 0.0111                  & 5.9322                  & 5.9322                   & 9.8300            & 9.4600            & 0.3700           & 3.7640                                                       \\ \midrule
149603524 & 3      & 4.4119    & 4.4285    & 0.0166   & 0.3763    & 0.0120                  & 0.0128                  & 6.6667                  & 6.6667                   & 10.8700           & 11.1500           & 0.2800           & 2.5759                                                       \\ \midrule
268766053 & 3      & 3.3100    & 3.3600    & 0.0500   & 1.5106    & 0.0177                  & 0.0169                  & 4.4604                  & 4.4604                   & 5.4800            & 5.6700            & 0.1900           & 3.4672                                                       \\ \midrule
388104525 & 4      & 2.4998    & 2.5340    & 0.0342   & 1.3681    & 0.0129                  & 0.0121                  & 5.9829                  & 5.9829                   & 7.3800            & 7.1654            & 0.2146           & 2.9079                                                       \\ \midrule
38846515  & 4      & 2.8493    & 2.8803    & 0.0310   & 1.0880    & 0.0069                  & 0.0068                  & 1.8813                  & 1.8813                   & 9.8200            & 9.9600            & 0.1400           & 1.4257                                                       \\ \midrule
289793076 & 1      & 3.0440    & 3.0548    & 0.0108   & 0.3548    & 0.0197                  & 0.0201                  & 2.2380                  & 2.2380                   & 9.2500            & 9.1300            & 0.1200           & 1.2973                                                       \\ \midrule
300871545 & 1      & 4.8170    & 4.8450    & 0.0280   & 0.5813    & 0.0236                  & 0.0253                  & 6.9915                  & 6.9915                   & 5.8100            & 5.7300            & 0.0800           & 1.3769                                                       \\ \midrule
231663901 & 1      & 1.4300    & 1.4292    & 0.0008   & 0.0559    & 0.0198                  & 0.0192                  & 3.0303                  & 3.0303                   & 5.4800            & 5.7800            & 0.3000           & 5.4745                                                       \\ \midrule
234523599 & 1      & 3.7960    & 3.7940    & 0.0020   & 0.0527    & 0.0484                  & 0.0462                  & 4.5455                  & 4.5455                   & 18.8400           & 18.6500           & 0.1900           & 1.0085                                                       \\ \midrule
290131778 & 1      & 3.3089    & 3.3167    & 0.0078   & 0.2357    & 0.0025                  & 0.0036                  & 44.0000                 & 44.0000                  & 4.3300            & 4.4250            & 0.0950           & 2.1940                                                       \\ \midrule
97409519  & 1      & 3.3729    & 3.3406    & 0.0323   & 0.9576    & 0.0160                  & 0.0138                  & 13.7500                 & 13.7500                  & 9.1100            & 9.1820            & 0.0720           & 0.7903                                                       \\ \midrule
281459670 & 1      & 3.1740    & 3.1857    & 0.0117   & 0.3698    & 0.0143                  & 0.0134                  & 6.2456                  & 6.2456                   & 8.7200            & 8.8100            & 0.0900           & 1.0321                                                       \\ \midrule
260609205 & 1      & 4.4620    & 4.4711    & 0.0091   & 0.2035    & 0.0171                  & 0.0179                  & 4.6784                  & 4.6784                   & 6.9200            & 6.8200            & 0.1000           & 1.4451                                                       \\ \midrule
25155319  & 1      & 3.2890    & 3.4147    & 0.1257   & 3.8214    & 0.0061                  & 0.0085                  & 38.4365                 & 38.4365                  & 7.8800            & 7.6500            & 0.2300           & 2.9188                                                       \\ \midrule
25375553  & 1      & 2.3110    & 2.3500    & 0.0390   & 1.6876    & 0.0076                  & 0.0080                  & 5.6150                  & 5.6150                   & 4.3500            & 4.5000            & 0.1500           & 3.4483                                                       \\ \bottomrule
\end{tabularx}
\label{tab: preds}
\end{table}
\finishlandscape

In addition, is important to remark that analyzing 150,000 light curves for each model took 30 s to complete, which is similar to three times the time required for analyzing 1 light curve with TLS in our server. TLS aims to maximize transit-like signal detection while decreasing the executing time as much as possible. However, as~least squares algorithms allow for choosing different prior parameters, the~amount of time required considerably depends on the density of the priors. For~example, the~minimum depth value considered during the analysis or the period intervals in which to search for periodic signals considerably constrain the execution time to complete the analysis. In~addition, the~light curves' length (in points) and their time span have also high impact. As~shown in~\cite{TLS}, the~executing times quadratically depend on the time span of the light curve. Obviously, the~computation power available is also fundamental for reducing the computing times. They show that on an Intel Core i7-7700K, a mean 4000-points-length K2~\cite{2014PASP..126..398H} light curve takes about 10 s. This process would probably not be possible with current algorithms in common computing facilities if the dataset that has to be analyzed would be composed of many thousands of stars (as the ones provided by current observing facilities), because~current algorithms are highly time-consuming, and also, they need a high computational cost, which is not always available.

Apart from reducing the computational cost and time consumption, our approach avoids the data preparation that is required from current MCMC and least squares methods. Almost all of them require obtaining some parameters related to the star, such as the limb darkening coefficients, something that could be carried out with different algorithms, which need stellar information, such as the effective temperature ($T_{eff}$), the~metallicity ($[M/H]$), and microturbulent velocity, among~others. 
Is not always easy to obtain these parameters from stellar databases or to compute them, especially if the analysis is carried out on a dataset composed of many thousands of stars, as~the ones provided by current observing facilities. However, our 1D CNN models allow for determining all these parameters in real time only by applying a simple normalization to the input data 
in order to invert the light curves and to set them between 0 and~1.


\section{Conclusions}\label{sec: conclusions}
In this research, we went one step further than our previous work. The~light curve simulator was generalized to TESS data, and~was also modified to obtain phase-folded light curves (in addition to the complete ones) with Gaussian noise, mimicking those expected for TESS data. As~light curve data can be understood as temporal series and also the predictions have high shape dependency, it was thought that 1D CNN techniques would be the most accurate ML techniques as they learn from the input data due to the convolutional filters and they perform properly with temporal series~data.

Our two models are similar, but~changing the last layer to adjust them to the required output values. The~first model works with complete light curves and thus predicts the value of the orbital period. The~second model works with phase-folded light curves and estimates the values $(R_p[R_\star])^2$ and $a[R_\star]$.

The training process was carried out with a dataset of 650,000 light curves, but splitting 30\% of them to validate the training. The~training histories show correct performance of both processes (training and validation), where both of them decrease among the epochs without the existence of 
a huge difference between them (that would suppose overfitting). After~training both models, another dataset composed of 150,000 light curves was used to test the results with a set different from the one used for training it (the test dataset). The~most visual way to analyze the accuracy of the prediction is by plotting the predicted values against the simulated ones. In~addition, some statistics that allow for taking into account the average error with which the CNN predicts the values were computed. More concretely, the~mean average error (MAE) and a linear regression $R^2$ coefficient were considered. For~the orbital period, $MAE= 0.015d$ and $R^2= 0.980$ were obtained (predictions made with model 1). For~$(R_p[R_\star])^2$, $MAE= 0.0003$ and $R^2=0.974$ were obtained (predictions made with model 2). For~$a[R_\star]$, $MAE= 0.9113$ and $R^2=0.971$ were obtained (predictions made with model 2). The~values of the modes from the histograms computed with the absolute percentage errors (APEs) of $P(d)$, $(R_p[R_\star])^2$, and $a[R_\star]$ are, respectively, 0.003, 0.69, and 0.38. Apart from that, a~set of 10,000 light curves from the test dataset was analyzed with TLS, showing that both models characterize planetary systems from their host star light curves with similar accuracy compared with current algorithms. In~addition, both models were tested on real data obtained of CTL light curves from MAST. The~results obtained for $P(d)$, $R_p[R_\star]$, and $a_[R_\star]$ ($0.0502 $, $0.0004$, and $0.20$, respectively, of~the MAEs; $1.87\%$, $7.16\%$, and $2.98\%$, respectively, of~the MAPEs; and $0.98$, $0.976$, and $0.985$, respectively, of~$R^2$), show that both models are able to characterize planetary systems from TESS light curves with high accuracy, which means not only that they are well trained and thus can be used for characterizing new exoplanets from TESS mission, but~also that the light curve simulator is able to mimic with high fidelity the light curves expected from this mission. In~addition, our models reduce the computing time and computational cost required for analyzing such large datasets as the ones available from current observing facilities; more concretely, our models take 30 s to complete the analysis of the test dataset (150,000 light curves), which is similar to three times the time required for analyzing a single light curve with TLS. Moreover, with~our models it is not necessary to set the priors in which to compute the main parameters, which also increases the time~consumption. 

In addition, we are not only in agreement with the fact that CNN in general and 1D CNN in particular are a good choice for analyzing light curves trying to search for transiting exoplanets (as previous research do~\cite{2018MNRAS.474..478P, Zucker_2018, Ansdell_2018, 2019MNRAS.488.5232C, Shallue_2018, gupta2023harnessing, Haider_2022, cuellar, Iglesias_2023}), but~also that CNN algorithms are able to characterize planetary systems with high accuracy in a short period of~time.



\vspace{6pt}
\authorcontributions{Research: S.I.Á.; coding: S.I.Á. and E.D.A.; writing: S.I.Á.; reviewing and editing: S.I.Á., E.D.A., M.L.S.R., J.R.R., S.P.F., and F.J.d.C.J; formal analysis: S.I.Á., S.P.F., and M.L.S.R.; manuscript structure: S.I.Á., E.D.A., and J.R.R. All authors have read and agreed to the published version of the~manuscript.}

\funding{This reseach was funded by Proyecto Plan Regional by FUNDACION PARA LA INVESTIGACION CIENTIFICA Y TECNICA FICYT, grant number SV-PA-21-AYUD/2021/51301, Plan Nacional by Ministerio de Ciencia, Innovación y Universidades, Spain, grant number MCIU-22-PID2021-127331NB-I00 and Plan Nacional by Ministerio de Ciencia, Innovación y Universidades, Spain, grant number~MCINN-23-PID2022-139198NB-I00.}


\acknowledgments{This research has made use of the NASA Exoplanet Archive, which is operated by the California Institute of Technology, under~contract with the National Aeronautics and Space Administration under the Exoplanet Exploration Program. This research has made use of the Exoplanet Follow-up Observation Program (ExoFOP; DOI: 10.26134/ExoFOP5) website, which is operated by the California Institute of Technology, under~contract with the National Aeronautics and Space Administration under the Exoplanet Exploration Program. We acknowledge the use of public TOI Release data from pipelines at the TESS Science Office and at the TESS Science Processing Operations Center. This paper includes data collected by the TESS mission, which are publicly available from the Mikulski Archive for Space Telescopes (MAST).}

\conflictsofinterest{The authors declare no conflicts of~interest.}

\begin{adjustwidth}{-\extralength}{0cm}

\reftitle{References}

\PublishersNote{}
\end{adjustwidth}
\end{document}